\documentclass{emulateapj}

\newcommand\holix{Ho IX X-1}%
\newcommand\holii{Ho II X-1}%
\newcommand\meo{M81 X-6}%
\newcommand\hbeta{$\rm H\beta$}%
\newcommand\halpha{$\rm H\alpha$}%
\newcommand\kms{$\rm km\;s^{\rm -1}$}%
\newcommand\ergs{$\rm erg\;s^{\rm -1}$}%
\newcommand\msol{$M_{\odot}$}%
\newcommand{\heii}{\ion{He}{2}}%
\newcommand{\heiiw}{\ion{He}{2}$\; \lambda$4686}%
\newcommand{\oiiiw}{\ion{O}{3}$\; \lambda$5007}%
\newcommand{\siiw}{\ion{S}{2}$\; \lambda$6717}%
\def\simgt{\lower.5ex\hbox{$\; \buildrel > \over \sim \;$}}
\def\simlt{\lower.5ex\hbox{$\; \buildrel < \over \sim \;$}}

\slugcomment{\today}

\shorttitle{\heii\ Nebulae Around Ultra-luminous X-ray Sources}
\shortauthors{Moon et al.}

\begin{document}

\title{Large Highly-Ionized Nebulae Around Ultra-luminous X-ray Sources}

\author{Dae-Sik~Moon\altaffilmark{1},
Fiona A. Harrison\altaffilmark{2},
S. Bradley Cenko\altaffilmark{3},
Jamil A. Shariff\altaffilmark{1}}

\altaffiltext{1}{Department of Astronomy and Astrophysics, University of Toronto, Toronto, ON M5S 3H4, Canada; moon@astro.utoronto.ca, shariff@astro.utoronto.ca}
\altaffiltext{2}{Space Radiation Laboratory, California Institute of Technology, MS 320-47, Pasadena, CA 91125; fiona@srl.caltech.edu}
\altaffiltext{3}{Department of Astronomy, University of California, Berkeley, CA 94720-3411, USA; cenko@astro.berkeley.edu}

\begin{abstract}
We present the results of deep optical spectroscopic observations using the LRIS spectrograph on the 
Keck I 10-m telescope of three
ultra-luminous X-ray sources (ULXs), \holix; \meo; and \holii. Our observations reveal the existence of large
(100--200 pc diameter) highly-ionized nebulae, identified  by diffuse \heiiw\ emission, surrounding these sources.
Our results are the first to find  highly-ionized nebulae of this extent, and the detection in all three objects indicates 
this may be a common feature of ULXs.   In addition to the extended emission,  \holix\ has an unresolved central
component containing about one-third of the total \heii\ flux, with a significant velocity dispersion of $\simeq$ 370 \kms,
suggestive of the existence of a photo-ionized accretion disk or an extremely hot early-type stellar counterpart.
Most of the \heii\ emission appears to be surrounded by significantly more extended \hbeta\ emission,
and the intensity ratios between the two lines, which range from 0.12 -- 0.33,  indicate that photo-ionization 
is the origin of the \heii\  emission.   Sustaining these extended nebulae requires substantial X-ray emission,
in the range $\sim 10^{39} - 10^{40}$~\ergs , comparable to the measured X-ray luminosities of the sources.
This favors models where the X-ray emission is isotropic, rather than beamed, which includes the interpretation
that ULXs harbor intermediate-mass black holes.
\end{abstract}


\section{Introduction}

Ultra-luminous X-ray sources (ULXs) are bright point-like sources found at off-nuclear positions in nearby
galaxies. If the X-ray emission is isotropic, the ULX luminosities would be in the range 10$^{39}$ -- 10$^{41}$ \ergs,
with corresponding  Eddington-limited masses  in excess of 100 \msol.  This leads to the suggestion that ULXs
are black holes with mass intermediate between stellar-mass systems and very massive nuclear black holes of galaxies.
The large luminosities, however, may still originate from stellar-mass black holes if the X-ray emission is beamed, 
or if accretion disks are capable of radiating at super-Eddington rates \citep[e.g.,][]{ket01, b02}.
Indeed some X-ray binaries are known to produce  super-Eddington luminosities,
reaching \simgt\ 10$^{41}$ \ergs\ in their flaring states \citep[e.g.,][]{me01, mew03a, mew03b, ssj05}.
The nature of ULXs is still quite uncertain,  and given they are broadly selected as bright off-nuclear
X-ray sources,  they may represent a mixture of different types of objects \citep{zr09}.

Many ULXs have early-type stars as optical counterparts, indicating their binary nature.
Some also have extended optical nebulae surrounding the central source \citep[see][for early results]{pm02}.
Of particular interest is the detection of the \heii\ line at 4686 \AA\ (\heii\ $\lambda$4686)
in the immediate vicinity (\simlt\ 2\arcsec) of some of bright sources 
\citep[e.g.,][]{get06, pm02, kwz04, let05, kc09}.
Because of the very high ionization potential of 54.4 eV, 
this line emission traces high-energy (EUV to X-ray) radiation fields or extremely strong shocks,
and is usually considered to be a signpost of X-ray photo-ionized nebulae  \citep[e.g.,][]{pa86, pm89}.
Where present, the observed \heiiw\ luminosities have been used to obtain
lower limits to the total X-ray ray luminosity  \citep[e.g.,][]{kc09},
which provides a constraint central to determining the nature of ULXs.

In this {\em Letter}, we present deep spectroscopic observations of the \heiiw\ line in three ULXs, \holix; \meo; and \holii.
\holix\ (= M81 X-9) is located in the dwarf galaxy Holmberg IX near M81,
and is known to have variable X-ray emission with a luminosity (assuming isotropic emission)
of $\sim$~10$^{40}$ \ergs\ \citep{lpet01},
along with a large ($\ge$ 250 pc in diameter) optical nebula seen in H recombination and forbidden lines \citep{m95, get06}.
Its optical counterpart, which is probably an early-type star of $V$ $\simeq$ 22.8 \citep{get06},
shows strong, broad, spatially unresolved \heiiw\ line emission \citep{get06}; however \citet{ret06} was unable to detect 
extended \heiiw\ emission.  \meo\ (= NGC 3031 X-11) resides in the spiral galaxy M81 and
also shows long-timescale variable X-ray emission with an isotropic-equivalent luminosity of $\sim$~2 $\times$ 10$^{39}$ \ergs.
Its optical counterpart is an O8~V star of $V$ $\simeq$ 23.9 situated in a spiral arm with high local stellar density
\citep{rw00, let02} and embedded in an \halpha\ nebula  $\sim$~150 pc in diameter \citep{ret06}.
Prior to the observations reported here no \heiiw\ emission has been detected in \meo.
\holii\ is one of the most luminous ($>$ 10$^{40}$ \ergs) ULXs,
and is located in the dwarf star-forming galaxy Holmberg II.
\citet{kwz04} identified a likely optical counterpart that is an early-type star
surrounded by a nearby optical nebula including \heiiw\ emission $\sim$~2\arcsec\ in size.
\citet{let05} confirmed the extended \heiiw\ region of 21 $\times$ 47 pc, and determined 
$B$ $\simeq$ 20.05 magnitude of the optical counterpart.  
In addition this ULX appears to have a radio counterpart \citep{mmn05}.
The observed \heiiw\ luminosity of 2.7 $\times$ 10$^{36}$ \ergs\ corresponds to a lower limit on the X-ray luminosity 
of 4--6 $\times$ 10$^{39}$ \ergs, assuming that X-ray photo-ionization is the source of the line excitation \citep{kwz04}.

\section{Observations and Results}

The observations of the three ULXs, \holix; \meo; and  \holii, occurred
2006 January 31, 2006 February 2--3, 
and 2007 March 3--5 using the Low Resolution Imaging Spectrograph \citep[LRIS;][]{oet95}
on the Keck~I 10-m telescope.  
The $B$-band seeing was 1.5\arcsec -- 2\arcsec\ (2006) and 1.0\arcsec -- 1.5\arcsec\ (2007);
the slit width and length were 1\arcsec\ and 120\arcsec, respectively.  
A beam dichroic inside LRIS splits incoming light to the blue and red sides followed by
a 600/4000 grism (3010--5600 \AA) and a 600/7500 grating (5700--8200 \AA).
Photometric calibrations were performed with standard sources HD~93521 and HD~9,
and wavelength solutions were obtained using arc lamp lines.

Table 1 summarizes the observations.   
We accumulated multiple exposures of 1800-s each for \holix\ and \meo, 
totalling on-source integration time up to 5.5 (\holix) and 7.0 (\meo) hours.
For \holii, which is much brighter,
we integrated for a total of $\sim$ 0.3 hours.
The slit position angles for \holix\ and \holii\ were fixed to be 0$^\circ$,
while two different position angles, 0$^\circ$ and 72.5$^\circ$, were used for \meo.
We detected extended \heii\ emission surrounding all three objects.  
The line was not detected in the  0$^\circ$ position angle observations of \meo , however
the signal to noise of these observations makes non-detection consistent with the level seen in the
2006 data. We therefore present the analyses of the 2006 data of \meo\ alone.
We also detected several Balmer series H lines and 
forbidden transitions of [\ion{O}{1}], [\ion{O}{2}], [\ion{O}{3}], [\ion{Ne}{3}], [\ion{N}{2}], and [\ion{S}{2}].
Detailed analyses of the entire spectra will be published elsewhere; in this {\it Letter} we concentrate
on the \heiiw\ and \hbeta\ line spatial and spectral analysis.

\begin{figure}[htf]
\plotone{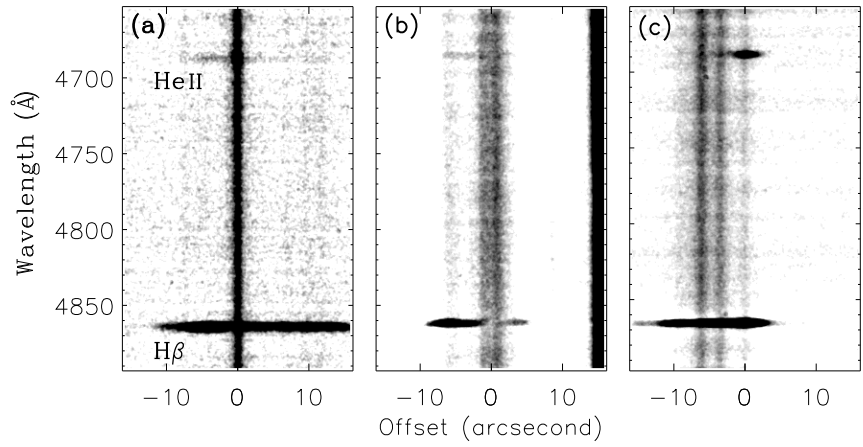}
\caption{
Integrated Keck LRIS spectrograms of the three ULXs -- \holix\ (a), \meo\ (b), and \holii\ (c) --
in the wavelength range of \heii\ ($\lambda$4686) and \hbeta\ ($\lambda$4861) lines.
The $x$-axis represents the angular distance (in the arcsecond unit)
from the optical counterparts of the ULXs.
Left is south for (a) and (c), whereas it is west-southwest for (b).
}
\label{fig_spectrogram}
\end{figure}

Figure~\ref{fig_spectrogram} shows integrated spectrograms of the three ULXs
in the wavelength range 4650 -- 4900 \AA,  containing the \heii\ ($\lambda$4686) 
and \hbeta\ ($\lambda$4861) lines. The offset ($x$-axis) is in arcseconds relative to the reported
position of the optical counterpart in each system.  The dispersed continuum emission from the 
optical counterparts of \holix\ and \holii\ is clearly visible in Figure~\ref{fig_spectrogram}(a) and (c), respectively.
In the \holii\ spectrogram, in addition to the optical counterpart,
there are two bright background stars between --3\arcsec\ and --7\arcsec. 
The counterpart detection in \meo\ in Figure~\ref{fig_spectrogram}(b) is uncertain.  
The dispersed light of its optical counterpart, which is fainter relative to the other two ULXs, 
appears to be blended with that of many other nearby stars \citep[see][]{let02}.
The $\simeq$ 2\arcsec\ seeing of these  observations make the continuum from the counterpart impossible to isolate from
that of nearby stars.    
(The bright continuum source at $\sim$ 15\arcsec\ to the right in Figure~\ref{fig_spectrogram}[b] is 
a field star used for the slit alignment.)    
In all three spectrograms extended \heii\ and \hbeta\ emission is clearly detected.

\begin{figure}[htf]
\plotone{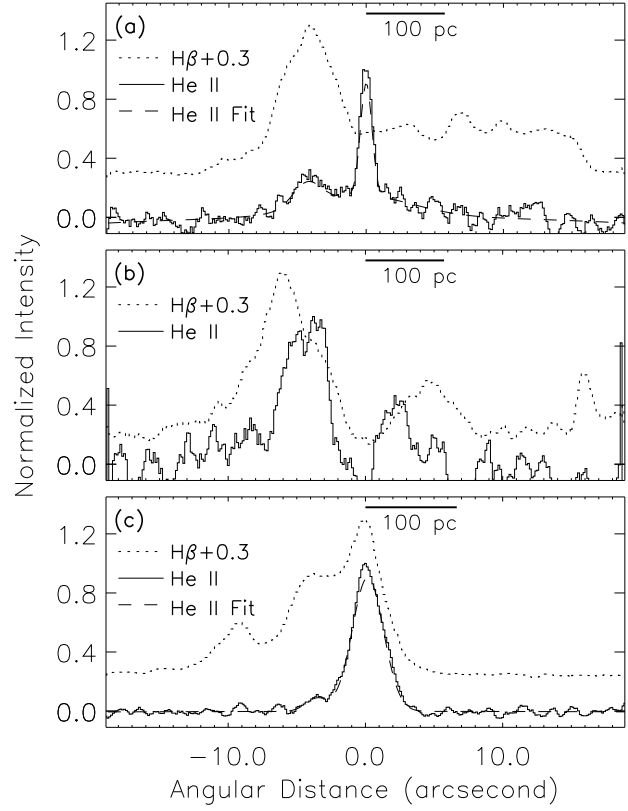}
\caption{
(a) Relative spatial distributions of the integrated \heii\ and \hbeta\ 
line intensities along the slit directions of \holix\ Figure~\ref{fig_spectrogram}
(solid line for the \heii\ line intensity distribution; dotted line for the \hbeta\ line),
along with the results of Gaussian fittings (dashed line) of the \heii\ line intensity distribution.
The solid bar at the top of the plot represents angular distance corresponding to 100 pc.
(b) Same as (a), but for \meo\ without Gaussian fitting results.
(c) Same as (a), but for \holii.
}
\label{fig_profile}
\end{figure}

Figure~\ref{fig_profile} compares the spatial distribution of the integrated 
\heii\ and \hbeta\ line intensities (solid line for \heii; dotted line for \hbeta)
along the slit direction, with the same $x$-axis origin used in Figure~\ref{fig_spectrogram}.  
The \hbeta\ distribution is shifted by an arbitrary constant (0.3) in the y-direction for clarity.
We subtracted the background, including the stellar continuum, using the flux detected in nearby pixels.
A common feature of all three (especially \holix\ and \holii) profiles is that, 
with the exception of the narrow central component in \holix, 
the \hbeta\ spatial profile generally tracks, 
but is more extended than, the \heii\ emission.
For all the three ULXs, the \heii\ emission extends to $>$100~pc from the center,
while the \hbeta\ emission extends much further, to beyond 250~pc.   
The \heiiw\ to \hbeta\ line intensity ratios are 
$\sim$0.12 (\holix), 0.19 (\meo), and 0.33 (\holii).
Below we discuss the spatial distribution for each source.

The \heii\ line intensity distribution in \holix\ (Figure~\ref{fig_profile}[a]) can be fit using three gaussian
components; the fit is shown as a dashed line.   The central, narrow, unresolved component is described by a gaussian profile with a 
FWHM of 1.0\arcsec\ (or 17.5 pc), comparable to the seeing size.   Its integrated intensity contains $\sim$1/3 of the total \heii\ emission.
This component is surrounded by asymmetric emission extending more than 100~pc in both directions, 
but more prominently to the south (left).   The extended \heii\ emission can be fit using two Gaussian components: 
one centered at the origin with FWHM of 8.7\arcsec\ ($\simeq$ 150 pc),
the other at --4.3\arcsec\ ($\simeq$ 75 pc) with FWHM of 2.4\arcsec\ ($\simeq$ 40 pc). 
The first component is roughly 3 times brighter than the second one,
and the intensity of the total emission in the south is greater than that in the north by $\sim$40\%.
We estimate the extent of the \heii\ emission to be $184 \pm 20$~pc, determined by
measuring the continuous area across the center where the signal intensity of each
data point is greater than noise level at the 90~\% confidence level.
We estimated the noise level at $<$ --15\arcsec\ and $>$ +15\arcsec,
where there is no apparent \heii\ emission.
The \hbeta\ emission of this source is asymmetric without any apparent central peak.   
With the notable exception
of the central peak, however, the \hbeta\ profile is similar in shape to that of the \heii.
The \hbeta\ spreads somewhat flatly to $\sim$300~pc in the northern direction, 
whereas, in the south, it extends less ($\sim$200~pc) and peaks
where the \heii\ emission has a local peak.  The total extension of the \hbeta\ emission, therefore, is $\sim$500 pc,
and it is slightly ($\sim$10~\% ) brighter in the south than the north.   
We estimate the total luminosity of the \heii\ line emission
integrated along the slit to be  $\sim$ 3.7 $\times$ 10$^{35}$ \ergs\ for a distance of 3.6 Mpc.

There is no identifiable central component to the \heii\ emission in \meo\  (Figure~\ref{fig_profile}[b]).
The distribution is irregular, and not well described by gaussian components.
The \heii\ is extended $115 \pm13$~pc west-southwest (left) from the center; 
whereas it extends $42 \pm 7$~pc in the opposite direction, for a total width of $167 \pm 15$~pc.
Both the \heii\ and \hbeta\ emission is heavily ($>$ 90~\%) concentrated in the west-southwest
where the \hbeta\ emission extends more than 150~pc.   
The \heii\ and \hbeta\ emission is again spatially correlated; however,
the peak locations of the \heii\ emission appear to be shifted toward the center compared to \hbeta.  
The main peak of the \heii\ emission may be dividable into two components at $x \simeq 4.5$\arcsec,
although the low signal-to-noise ratio makes it difficult to confirm this.
The total measured \heii\ luminosity is $\sim 8.8 \times 10^{34}$~\ergs\ for a distance of 3.6 Mpc.

For \holii\ (Figure~\ref{fig_profile}[c]), the \heii\ emission is well-described
by a broad, central Gaussian component of 2.7\arcsec\ ($\simeq$ 40 pc) FWHM,
along with a minor component of 1.9\arcsec\ ($\simeq$ 30 pc) FWHM
located at --3\farcs5 ($\simeq$ 50 pc) to the south.
The integrated intensity of the former is roughly 10 times greater than that of the latter.
The \hbeta\ emission is distributed very similarly to the \heii\ emission:
the main peak is at the center and there is a secondary peak near the location 
of the secondary component of the \heii\ emission at --3\farcs5. 
In addition, there is a third peak centered at --10\arcsec\ in the \hbeta\ emission. 
The \heii\ emission is somewhat enhanced at this location, 
although it is almost indistinguishable from the noise. 
The size of the \heii\ emission is 122 $\pm$ 7 pc.
The measured \heii\ luminosity is $\sim$ 3.6 $\times$ 10$^{36}$ \ergs,
which is slightly greater than that of previous measurements \citep{pm02, kwz04, let05},
for a distance of 3.1 Mpc.  

\begin{figure}[htf]
\plotone{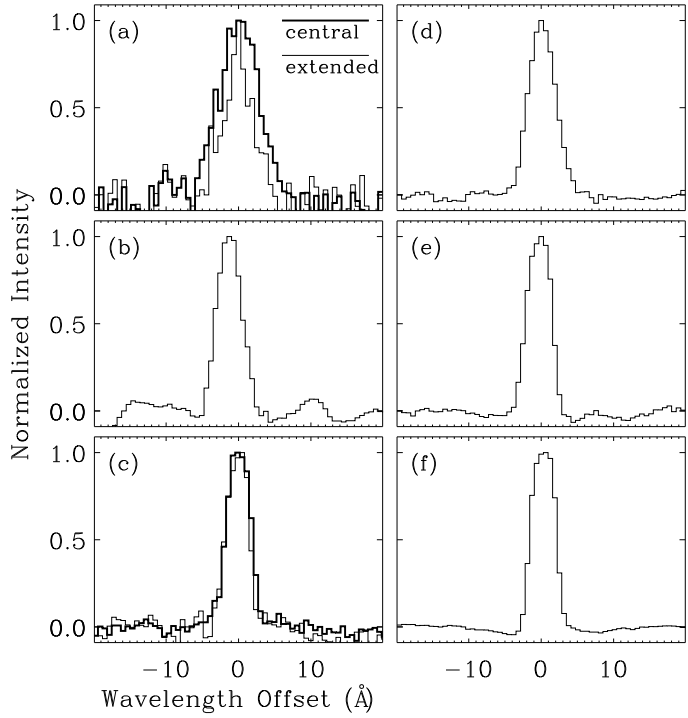}
\caption{
({\it Left}) (a) \heiiw\ line profiles for the central (thick solid line) and extended emission (thin solid line)
of \holix.
(b) Same as (a), but for the extended emission of \meo.
(c) Same as (a), but for \holii.
({\it Right}) (d) \hbeta\ line profile of \holix.
(e) Same as (d), but for \meo.
(f) Same as (d), but for \holii.
}
\label{fig_line}
\end{figure}

Figure~\ref{fig_line} shows spectral line profiles of the \heii\ (left panels)
and \hbeta\ (right panels) transitions for the three ULXs.  There are two \heii\ line profiles for \holix\ and \holii:
the thick-solid profiles are for the central emission obtained within 1\arcsec\ from the center 
(i.e., the optical counterparts); the thin-solid ones are for the extended emission outside the center.
The FWHM of the \heii\ lines of the spatially extended emission are  
5.0 (\holix), 3.7 (\meo), and 3.5 (\holii) \AA.
Those for the central emission are 6.7 \AA\ (\holix) and 3.6 \AA\ (\holii).
For the \hbeta\ lines,  the line widths of the extended emission are 
4.4 \AA\ (\holix), 3.6 \AA\ (\meo), and 3.5 \AA\ (\holii).  
The LRIS instrumental line widths, estimated from
the measured widths of the calibration lamp lines,
are in the range  3.8 -- 4.1 \AA.
We therefore conclude that the \heii\ emission of \holix, especially the central component, 
has a contribution from dynamical motion in the source,
whereas we can only determine an upper limit of $\sim$ 250 \kms\ for the velocity dispersion of the other lines.
For \holix, the extra widths correspond to velocity dispersions of 
370 \kms\ and 230 \kms\ for the central and extended components, respectively.  
These dispersions of the \heii\ lines are consistent with previous observations \citep{get06},
and are larger than those reported in \oiiiw\ and \siiw\ lines \citep{am08}.
For \holii, the upper limit is consistent with the results of previous observations \citep{let05}.

\section{Discussion and Conclusions}

Our deep Keck observations spectroscopically identified 
spatially extended, highly-ionized \heii\ emission around  three ULXs.   
In the case of \holix\ and \meo\ this is the first reported
detection of  diffuse \heii\ emission from these systems, and in the case of \holii, we find the highly-ionized region to be larger than
previously reported.    The sizes we find are in the range of 100 -- 200 pc (in diameter),  larger 
than any previously-known \heii\ emission around a compact source.  The \heii\ nebulae around extremely hot stellar sources,
such as planetary nebulae or stellar X-ray sources,
are generally smaller than 10 pc.
Previous optical observations of some ULXs, including \holix\ and \holii\ observed in this study, detected \heii\ emission only 
from the locations of the optical counterparts \citep[e.g., M101 X-1, NGC 1313 X-1, and \holix;][]{ket05, pet05, get06}
or from their close ($<$ 50 pc) vicinities  \citep[e.g., \holii\ and NGC 5408 X-1;][]{get06, kc09}.

The identification of  large \heii\ emission regions surrounding all three sources included in this study indicates
that extended, highly ionized nebulae are a common or even ubiquitous feature of ULXs.  Previous non-detections
likely result from the limited depth of the observations; with Keck LRIS being significantly more sensitive than
other telescopes for this purpose.  In all three cases the
extended emission contains the majority of the total  \heii\ flux.  In the case of \holix\ the central unresolved component
amounts to only 1/3 of the total emission,  and there is no apparent central component at all in \meo.  
In \holii\  the \heii\ emission is distributed in a relatively broad ($\sim$ 40 pc FWHM), but very regularly distributed
component with a minor contribution from very extended emission in the south.   

A distinct feature of the \heii\ emission in \holix\ is the unresolved central component of a significant
($\simeq$ 370 \kms) velocity dispersion without any apparent \hbeta\ counterpart. 
One possible explanation is that the central component represents a photo-ionized accretion disc rotating 
around the central X-ray source \citep{get06}. 
The line broadening in this case is due to rotational motion, 
and the lack of corresponding \hbeta\ emission suggests that it is gas of relatively small column density.   
The large velocity dispersion implies that the disk size 
is much smaller than the seeing size for any conceivable mass range for the central source, consistent with our results (\S~2).
As noted above, \meo\ lacks a central component altogether.  
In \holii\  the absence of a large velocity dispersion in the central region 
(offset of $\leq$ 1\arcsec) probably indicates the lack of an accretion disc, 
although it may be due to a projection effect, i.e., \holii\ is close to a face-on system.   
The more likely scenario is that the  extended 
highly ionized nebula is regularly distributed around the system's center and smoothly connects to the broad component.
Another possibility of the unresolved central \heii\ component of \holix\ is that it is due
to strong \heii\ emission from the optical (stellar) counterpart of the source.
For instance, some Wolf-Rayet stars are known to produce luminous \heii\ emission of significant 
velocity dispersion \citep[e.g.,][]{ch06}. 
We need more information on the optical counterpart to investigate this scenario further.

Our observations point to photo-ionization by intense X-ray flux being the source of the \heii\ and \hbeta\ emission.
The alternative explanation is that both are produced by the same radiative shocks.  
However, in the case of \meo\ and \holii\ the observed line intensity ratios of \heii\ to \hbeta\
require much greater velocities than the measured upper limits \citep{aet08},
making it incompatible with radiative shocks.
For \holix, the observed line intensity ratio requires comparable,
but still slightly greater, velocities than observed ($\sim$230 \kms\ for \heii) if the emission is produced in strong shocks.
Therefore, our results favor the interpretation that  the \heii\ line emission is dominated by X-ray photo-ionization.
Under this scenario, the smaller extent of the \heii\ emission traces  the locations where most of the energetic X-ray 
radiation from the ULXs is locally absorbed.

Under the assumption of photo-ionization, the observed \heii\ luminosities can be used to obtain independent measurements
of the ULX X-ray luminosities. The X-ray luminosities estimated based on the observed \heii\ line luminosities
are independent of the assumption that the X-ray radiation from the central source is isotropic
-- the inevitable assumption to calculate the X-ray luminosities from the observed X-ray fluxes.
Measurements to-date based on the line emission all indicate luminosities
in excess of 10$^{39}$ \ergs.  For example, the observed \heii\ line luminosity of $\sim$1 $\times$ 10$^{36}$ \ergs\ 
of the ULX in NGC 5408 provides a lower limit of $\sim$2.5 $\times$ 10$^{39}$ \ergs\ 
for its X-ray luminosity \citep{kc09} and
the \heii\ luminosity of $\sim$2.7 $\times$ 10$^{36}$ \ergs\ leads to an estimated X-ray luminosity
of $\sim$5 $\times$ 10$^{39}$ \ergs\ in \holii.

In our observations we use a relatively small (1\arcsec\ in width)  slit, so that the measured luminosities given in 
\S~2 are firm lower limits. 
One way to estimate the upper limits on the extended \heii\ line luminosities is
to scale up the observed luminosities within the slit under the hypothesis that 
the \heii\ emission is uniformly and symmetrically distributed outside the slit.
In doing so we obtain the upper limits of $\sim$1.2 $\times$ 10$^{37}$ \ergs\ and 
$\sim$4.4 $\times$ 10$^{36}$ \ergs\ for \holix\ and \meo, respectively,
within an 8\arcsec\ radius. 
We then calculate the X-ray luminosities required to produce the \heii\ line luminosities
using the photo-ionization modelling program $CLOUDY$ \citep[][ver. 07.02.01]{fet98}.
In the modelling, we used an input X-ray spectrum of a multicolor disc blackbody
plus a cutoff power law component
for \holix\ \citep{dgr06}, and disk blackbody component for \meo\ \citep{set03}.
We also assumed a constant gas density of 10 cm$^{-3}$, a unity filling factor,
and a spherically symmetric geometry.
We find that to account for the \heii\ line luminosities the required X-ray luminosities are 
in the range of $\sim$10$^{39}$ -- 10$^{40}$ \ergs,
comparable to the observed ULX X-ray luminosities. 
This favors models where the X-ray emission is isotropic rather than beamed, 
which includes the scenario where ULXs are intermediate mass black holes.

\acknowledgments
D.-S.M. acknowledges the support by NSERC through Discovery program 327277 and
S.B.C. acknowledges generous support from Gary and Cynthia Bengier and the Richard and Rhoda Goldman fund.

The data presented herein were obtained at the W.M. Keck Observatory, which is operated as a scientific partnership among
the California Institute of Technology, the University of California, and National Aeronautics and Space Administration.  The
Observatory was made possible by the generous financial support of the W.M. Keck Foundation.

{\em Facilities:} Keck:I (LRIS)

\clearpage
\begin{deluxetable}{lccll}
\tablecolumns{5}
\tablewidth{0pt}
\tablecaption{Observing Parameters}
\tablehead{
\colhead{Object} & \colhead{Coordinate (J2000)} & \colhead{Position Angle} & \colhead{Date} & \colhead{Exposure} }
\startdata
\holix & (09:57:53.3, +69:03:48) & 0    & 2007 Mar. 24--26        & 1800 s $\times$ 11  \\
\meo   & (09:55:33.0, +69:01:13) & 72.5   & 2006 Jan. 31, Feb. 2--3 & 1800 s $\times$ 9  \\
\meo   & (09:55:33.0, +69:01:13) & 0    & 2007 Mar. 24--25        & 1800 s $\times$ 5  \\
\holii & (08:19:29.0, +70:42:19) & 0    & 2007 Mach 26            &  600 s $\times$ 2  \\
\enddata
\end{deluxetable}

\clearpage

\clearpage

\clearpage


\begin{thebibliography}{}
\bibitem[Abolmasov \& Moiseev(2008)]{am08}
    Abolmasov, P., \& Moiseev, A. V. 2008, RMxAA, 44, 301
\bibitem[Allen et al.(2008)]{aet08}
    Allen, M. G., Groves, B. A., Dopita, M. A., Sutherland, R. S., \& Kewley, L. J. 2008, \apjs, 178, 20
\bibitem[Begelman(2002)]{b02}
    Begelman, M. C. 2002, \apj, 568, L97
\bibitem[Crowther \& Hadfield(2006)]{ch06}
    Crowther, P. A., \& Hadfield, L. J. \aa, 449, 711
\bibitem[Ferland et al.(1998)]{fet98}
    Ferland, G. J., Korista, K. T., Verner, D. A., Ferguson, J. W., Kingdon, J. B., \& Verner, E. M. 1998, PASP, 110, 761
\bibitem[Grise et al.(2006)]{get06}
    Grise, F., Pakull, M. W., \& Motch, C. 2006, in IAU Symp. 230, Populations of High Energy Sources in Galaxies, ed. E. J. A.Meurs \& G. Fabbiano (Cambridge: Cambridge Univ. Press), 302 
\bibitem[Kaaret \& Corbel(2009)]{kc09}
    Kaaret, P., \& Corbel, S. 2009, \apj, 697, 950
\bibitem[Dewangan et al.(2006)]{dgr06}
    Dewangan, G. C., Griffiths, R. E., \& Rao, A. O. 2006, \apj, 641, L125
\bibitem[Kaaret et al.(2004)]{kwz04}
    Kaaret, P., Ward, M. J., \& Zezas, C. 2004, \mnras, 351, L83
\bibitem[King et al.(2001)]{ket01}
    King, A. R., Davies, M. B., Ward, M. J., Fabbiano, G., \& Elvis, M. 2001, \apj, 552, L109
\bibitem[Kuntz et al.(2005)]{ket05}
    Kuntz, K. D., Gruendl, R. A., Chu, Y.-H., Chen, C.-H. R., Still, M., Mukai, K., \& Mushotzky, R. F. 2005, \apj, 620, L31
\bibitem[La Parola et al.(2001)]{lpet01}
    La Parola, V., Peres, G., Fabbiano, G., Kim, D. W., \& Bocchino, F. 2001, \apj, 556, 47
\bibitem[Lehmann et al.(2005)]{let05}
    Lehmann, I., et al. 2005, A\&A, 431, 847
\bibitem[Liu et al.(2002)]{let02}
    Liu, J.-F., Bregman, J. N., \& Seitzer, P. 2002, \apj, 580, L31
\bibitem[Miller(1995)]{m95}
    Miller, B. W. 1995, \apj, 446, L75
\bibitem[Miller et al.(2005)]{mmn05}
    Miller, N. A., Mushotzky, R. F., \& Neff, S. G. 2005, \apj, 623, L109
\bibitem[Moon \& Eikenberry(2001)]{me01}
    Moon, D.-S., \& Eikenberry, S. S. 2001, \apjl, 549, L225
\bibitem[Moon et al.(2003a)]{mew03a}
    Moon, D.-S., Eikenberry, S. S., \& Wasserman, S. M. 2003a, \apjl, 582, L91
\bibitem[Moon et al.(2003b)]{mew03b}
    ------. 2003b, \apj, 586, 1280
\bibitem[Oke et al.(1995)]{oet95}
    Oke, J. B., et al. 1995, \pasp, 107, 375
\bibitem[Pakull \& Angebault(1986)]{pa86}
    Pakull, M. W., \& Angebault, L. P. 1986, \nat, 332, 511
\bibitem[Pakull et al.(2005)]{pet05}
    Pakull, M. W., Grise, F., \& Motch, C. 2006, in Proc. IAU Symp. 230, Populations
    of High Energy Sources in Galaxies, ed. E. J. A. Meurs \& G.
    Fabbiano (Cambridge: Cambridge Univ. Press), 293
\bibitem[Pakull \& Mirioni(2002)]{pm02}
    Pakull, M. W., \& Mirioni, L. 2002, preprint (astro-ph/0202488)
\bibitem[Pakull \& Motch(1989)]{pm89}
    Pakull, M. W., \& Motch, C. 1989, \nat, 337, 337
\bibitem[Ramsey et al.(2006)]{ret06}
    Ramsey, C. J., Williams, R. M., Gruendl, R. A., Chen, C.-H. R., Chu, Y.-H., \& Wang, Q. D. 2006, \apj, 641, 241
\bibitem[Roberts \& Warwick(2000)]{rw00}
    Roberts, T. P., \& Warwick, R. S. 2000, \mnras, 315, 98
\bibitem[Sivakoff et al.(2005)]{ssj05}
    Sivakoff, G. R., Sarazin, C. L., \& Jord\'{a}n, A. \apj, 624, L17
\bibitem[Swartz et al.(2003)]{set03}
    Swartz, D. A., Ghosh, K. K., McCollough, M. L., Pannuti, T. G., Tennant, A. F. \& Wu, K. 2003, \apjs, 144, 213
\bibitem[Zampieri \& Roberts(2009)]{zr09}
    Zmpieri, L., \& Roberts, T. P. 2009, \mnras, 400, 677
\end{thebibliography}
\end{document}